\documentclass{aa}       
\usepackage{psfig}

{\catcode`\@=11 \gdef\SchlangeUnter#1#2{\lower2pt\vbox{\baselineskip 0pt
\lineskip0pt
  \ialign{$\m@th#1\hfil##\hfil$\crcr#2\crcr\sim\crcr}}}
}
\def\gtrsim{\mathrel{\mathpalette\SchlangeUnter>}}
\def\lessim{\mathrel{\mathpalette\SchlangeUnter<}}

\renewcommand\deg{\ifmmode^\circ\else$^\circ$\fi}

\begin{document}

\thesaurus{03         
              (11.01.2;
               11.02.2 AO\,0235+164;
               13.18.1)
             }

\title{Unusual radio variability in the BL\,Lac object 0235+164}

\author{A. Kraus\inst{1} \and A. Quirrenbach\inst{2,3} \and
        A.P. Lobanov\inst{1} \and T.P. Krichbaum\inst{1} \and
        M. Risse\inst{1} \and P. Schneider\inst{4} \and
        S.J. Qian\inst{1,7} \and
        S.J. Wagner\inst{5} \and A. Witzel\inst{1} \and
        J.A. Zensus\inst{1} \and
        J.\ Heidt\inst{5} \and  H.\ Bock\inst{5}  \and
        M. Aller\inst{6} \and H. Aller\inst{6} }

\offprints{A.~Witzel}

\institute{Max-Planck-Institut f\"{u}r Radioastronomie,
Auf dem H\"{u}gel 69,
53121 Bonn, Germany
\and
Max-Planck-Institut f\"ur Extraterrestrische Physik,
Giessenbachstr., Postfach 1603,
85740 Garching, Germany
\and
University of California San Diego,
Dept. of Physics,
Center for Astrophysics and Space Sciences,
Mail Code 0424,
La Jolla, CA 92093-0424,
USA (present address)
\and
Max-Planck-Institut f\"ur Astrophysik,
Karl-Schwarzschild-Str.~1,
85740 Garching, Germany
\and
Landessternwarte Heidelberg,
K\"onigstuhl,
69117 Heidelberg, Germany
\and
Astronomy Department, University of Michigan,
830 Dennison Building,
Ann Arbor, MI 48109-1090, USA
\and
Beijing Astronomical Observatory,
Chinese Academy of Science,
Beijing 100080, China
}

\date{received;accepted}

\maketitle
\markboth{A. Kraus et al.: Unusual radio variability of 0235+164}{}

\begin{abstract}

We present radio observations at three frequencies and contemporaneous
optical monitoring of the peculiar BL\,Lac object AO\,0235+164. During
a three-week campaign with the VLA we observed intraday variability in
this source and found a distinct peak which can be identified
throughout the radio frequencies and tentatively connected to the
R-band variations. This event is characterized by unusual properties:
its strength increases, and its duration decreases with wavelength,
and it peaks earlier at 20\,cm than at 3.6 and 6\,cm. We discuss
several generic models (a ``standard'' shock-in-jet model, a
precessing beam, free-free-absorption in a foreground screen,
interstellar scattering, and gravitational microlensing), and explore
whether they can account for our observations. Most attempts at
explaining the data on 0235+164 require an extremely small source
size, which can be reconciled with the $10^{12}$\,K inverse Compton
limit only when the Doppler factor of the bulk flow is of order
100. However, none of the models is completely satisfactory, and we
suggest that the observed variability is due to a superposition of
intrinsic and propagation effects.

\keywords{Galaxies: active --
BL\,Lacertae objects: individual: AO\,0235+164 -- Radio continuum: galaxies
               }
\end{abstract}

\section{Introduction}

The radio source AO\,0235+164 was identified by Spinrad \& Smith
(\cite{spinrad75}) as a BL\,Lac object due to its almost featureless
optical spectrum at the time of the observation, and due to its
pronounced variability.  Long-term flux density monitoring in the
radio and optical regimes have revealed strong variations and repeated
outbursts with large amplitudes and timescales ranging from years down
to weeks (e.g.\ Chu et al.\ \cite{chu96}, O'Dell et al.\
\cite{odell88}, Ter\"asranta et al.\ \cite{teraesranta92}, Schramm et
al.\ \cite{schramm94}, Webb et al.\ \cite{webb88}, this paper, Fig.\
\ref{fg:longplot}). Furthermore, intraday variability in the radio
(Quirrenbach et al.\ \cite{quirrenbach92}, Romero et al.\
\cite{romero97}), in the IR (Takalo et al.\ \cite{takalo92}), and in
the optical regime (Heidt \& Wagner \cite{heidt96}, Rabbette et al.\
\cite{rabbette96}) has also been observed in this object. In the high
energy regime, 0235+164 was detected with EGRET on board of the CGRO
(v.~Montigny et al.\ \cite{montigny95}), showing variability between
the individual observations. Madejski et al.\ (\cite{madejski96})
report variability by a factor of 2 in the soft X-rays during a ROSAT
PSPC observation in 1993. VLBI observations (e.g.\ Shen et al.\
\cite{shen97}, Chu et al.\ \cite{chu96}, B\aa{}\aa{}th \cite{baath84},
Jones et al.\ \cite{jones84}) reveal a very compact structure and
superluminal motion with extremely high apparent velocities (perhaps
up to $\beta_{\rm app} \simeq 30$).

Three distinct redshifts have been measured towards 0235+164 (e.g.\
Cohen et al.\ \cite{cohen87}). Whereas the emission lines at $z=0.940$
have been attributed to the object itself, two additional systems are
present in absorption ($z=0.851$) and in emission and absorption
($z=0.524$). Smith et al.\ (\cite{smith77}) observed a faint object
located about $2''$ south of 0235+164, and measured narrow emission
lines at a redshift of $z=0.524$. Continued studies on the field of
0235+164 have revealed a number of faint galaxies, mostly at a
redshift of $z=0.524$, including an object located 1\farcs 3 to the
east and 0\farcs 5 to the south (e.g.\ Stickel et al.\
\cite{stickel88}, Yanny et al.\ \cite{yanny89}). Recently, Nilsson et
al.\ (\cite{nilsson96}) investigated 0235+164 during a faint state and
found prominent hydrogen lines at the object redshift of
$z=0.940$. They note that 0235+164 -- at least when in a faint state
-- shows the spectral characteristics of an HPQ. Furthermore, through
HST observations of 0235+164 and its surrounding field, Burbidge et
al.\ (\cite{burbidge96}) discovered about 30 faint objects around
0235+164 and broad QSO-absorption lines in the southern companion,
indicating that the latter is an AGN-type object. Due to the presence
of several foreground objects, gravitational microlensing might play a
role in the characteristics of the variability in 0235+164, as was
suggested by Abraham et al.\ (\cite{abraham93}).

In this paper, we further investigate the radio variability of
0235+164, and attempt to determine the most likely physical mechanisms
behind the observed flux density variations. The plan of the paper is
as follows.  In Section 2 we describe the observations and the data
reduction; subsequently we analyze the lightcurves and point out some
of their special properties. In Section 3 we explore different
scenarios which could explain the variability: we discuss relativistic
shocks, a precessing beam model, free-free-absorption, interstellar
scattering, and gravitational microlensing. Finally, in Section 4
we conclude with a summary of the failures and successes of these
models.

Throughout the paper we assume a cosmological interpretation of the
redshift, we use $H_0 = 100 h$\,km/(s\,Mpc) and $q_0 = 0.5$, which
gives for the redshift of 0235+164 a luminosity distance of
3280\,$h^{-1}$\,Mpc; 1\,mas corresponds to 4.2\,$h^{-1}$\,pc. The
radio spectral index is defined by $S_{\nu} \propto \nu^{\alpha}$.

\section{Observations and data reduction}

\subsection{Radio observations}

From Oct 2 to Oct 23, 1992, we observed 0235+164 with a five-antenna
subarray of the VLA\footnote{The Very Large Array, New Mexico, is
operated by Associated Universities, Inc., under contract with the
National Science Foundation.}  during and after a reconfiguration of
the array from D to A. The aim of these observations was to search for
short-timescale variations in several sources.  The complete data set
will be presented elsewhere (Kraus et al., in preparation).  In
parallel, optical observations were performed in the R-band (see
Section \ref{optical}). Data for 0235+164 were taken at 1.49, 4.86,
and 8.44\,GHz ($\lambda =$~20, 6, 3.6\,cm) every two hours around
transit, i.e., six times per day.  These three sets of receivers have
the lowest system temperatures and highest aperture efficiencies of
those available at the VLA (see Crane \& Napier \cite{crane89}); and
data in these bands are less susceptible to problems with poor
tropospheric phase stability than those at higher frequencies. In
addition, intraday variability of compact flat-spectrum radio sources
did appear most markedly in this frequency range in previous
observations (Quirrenbach et al.\ \cite{quirrenbach92}). During the
first week, the antennae included in our subarray were changed
repeatedly due to the ongoing reconfiguration; however, an attempt was
made to maintain an approximately constant set of baselines.  Since
0235+164 and the used calibrator sources used are extremely compact
(cf.\ VLA calibrator list), the effect of the ongoing reconfiguration
on the measurements is negligible.

After correlation and elimination of erroneous data intervals, we
performed phase calibration first.  Subsequently, a (one-day) mean
amplitude gain factor was derived using non-variable sources such as
1311+678, which have been linked to an absolute flux density scale
(Baars et al.~1977, Ott et al.~1994) by frequent observations of
3C\,286 and 3C\,48.  After a second pass of editing spurious sections
of the data, the visibilities of each scan were incoherently averaged
over time, baselines, polarization, and IFs. Because of the point-like
structure of the sources, the mean source visibility is proportional
to the flux density.  Eventually, systematic elevation and
time-dependent effects in the lightcurves were removed, using
polynomial corrections derived from observations of the calibrator
sources 0836+710 and 1311+678.

The errors are composed of the statistical errors from the averaging
and a contribution from the residual fluctuations of the non-variable
sources 3C\,286, 1311+678 and 0836+710. The level of these
fluctuations was estimated from a running standard deviation of the
calibrator measurements over a two-day period. Over the full
three-week period, the standard deviations were found to be 0.5, 0.5,
and 0.7\,\% of the mean value at 1.5, 4.9, and 8.4\,GHz respectively
(with no significant difference between the three non-variable
sources).

\begin{figure}
\psfig{figure=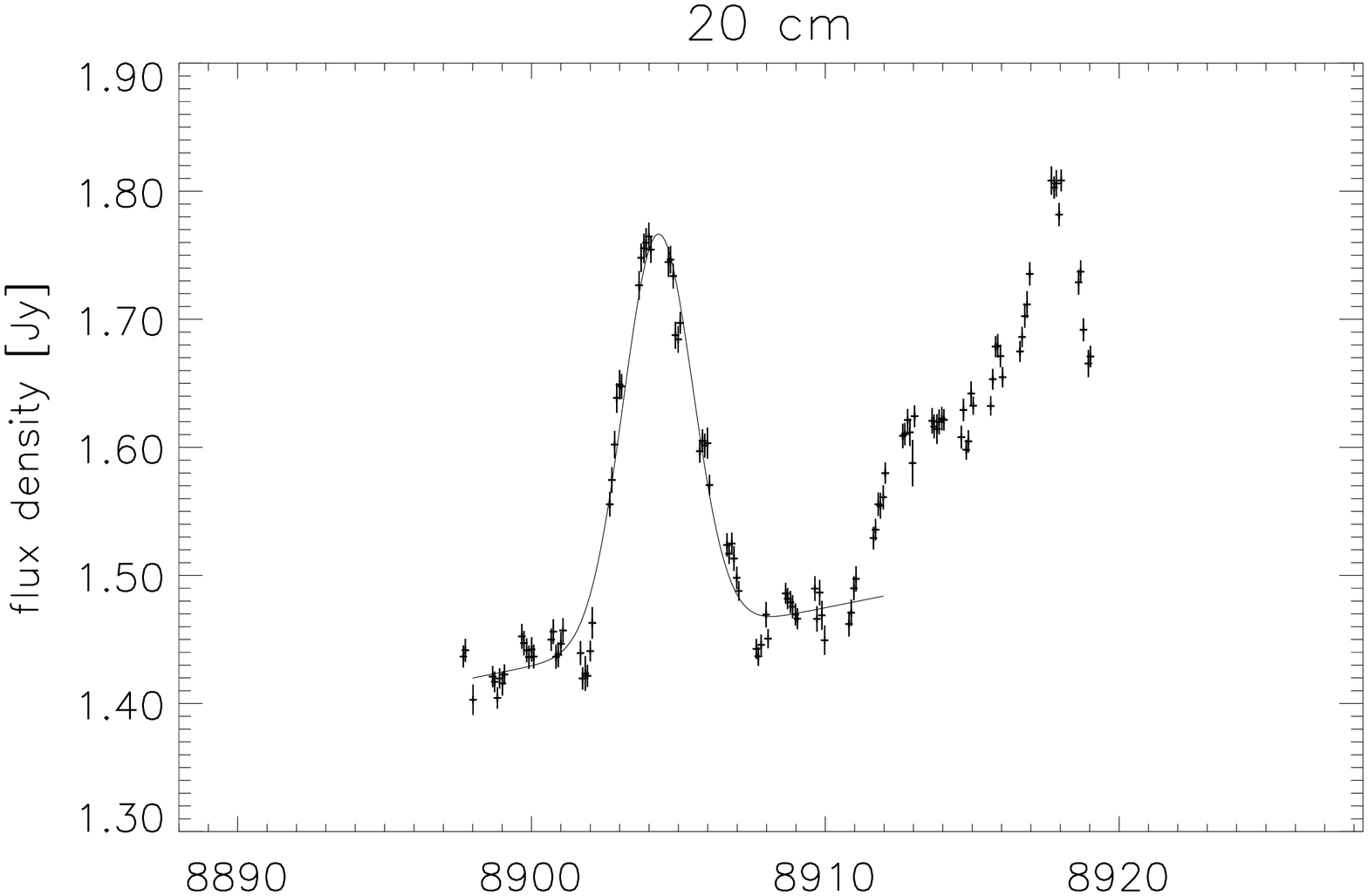,height=5cm,angle=90}
\psfig{figure=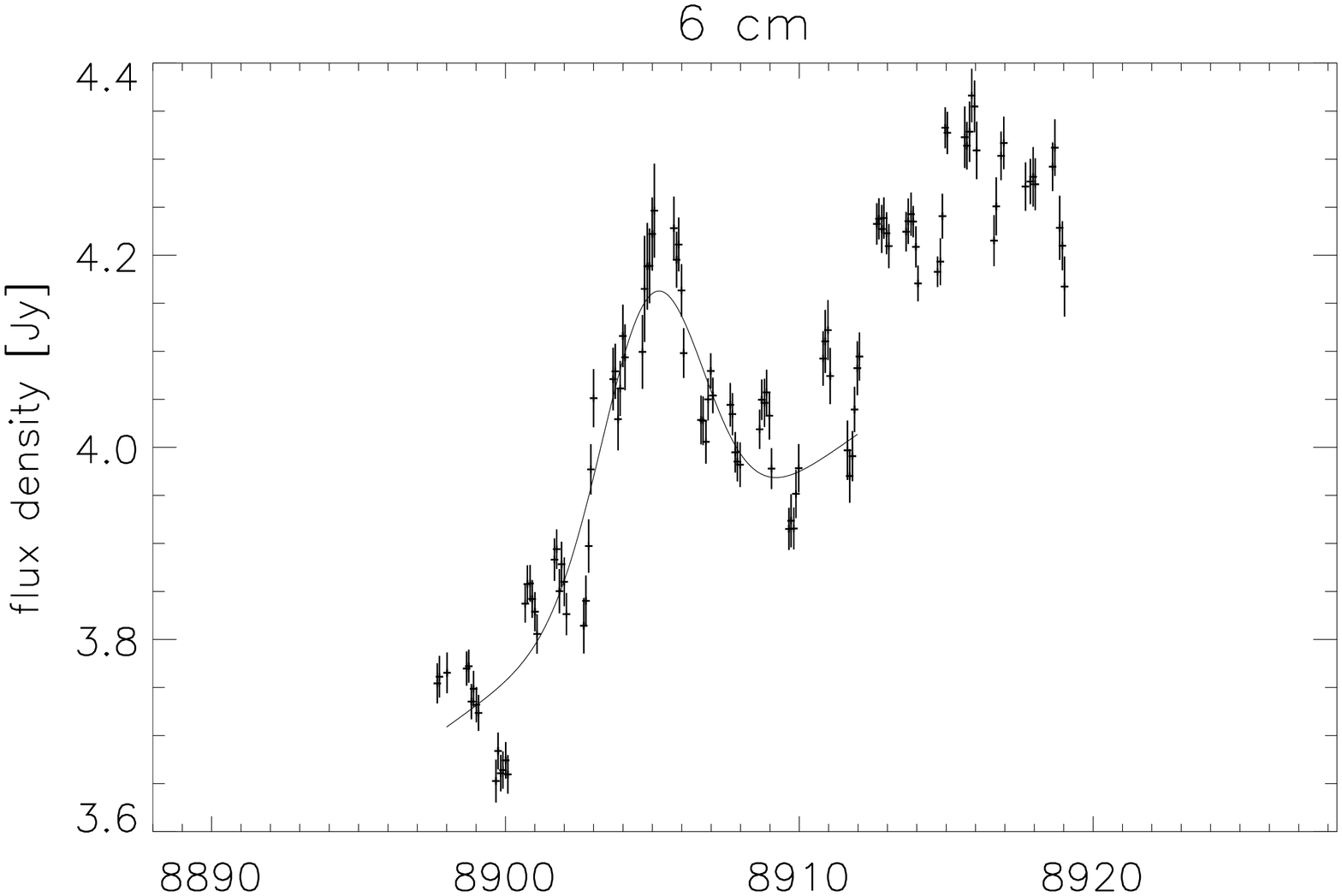,height=5cm,angle=90}
\psfig{figure=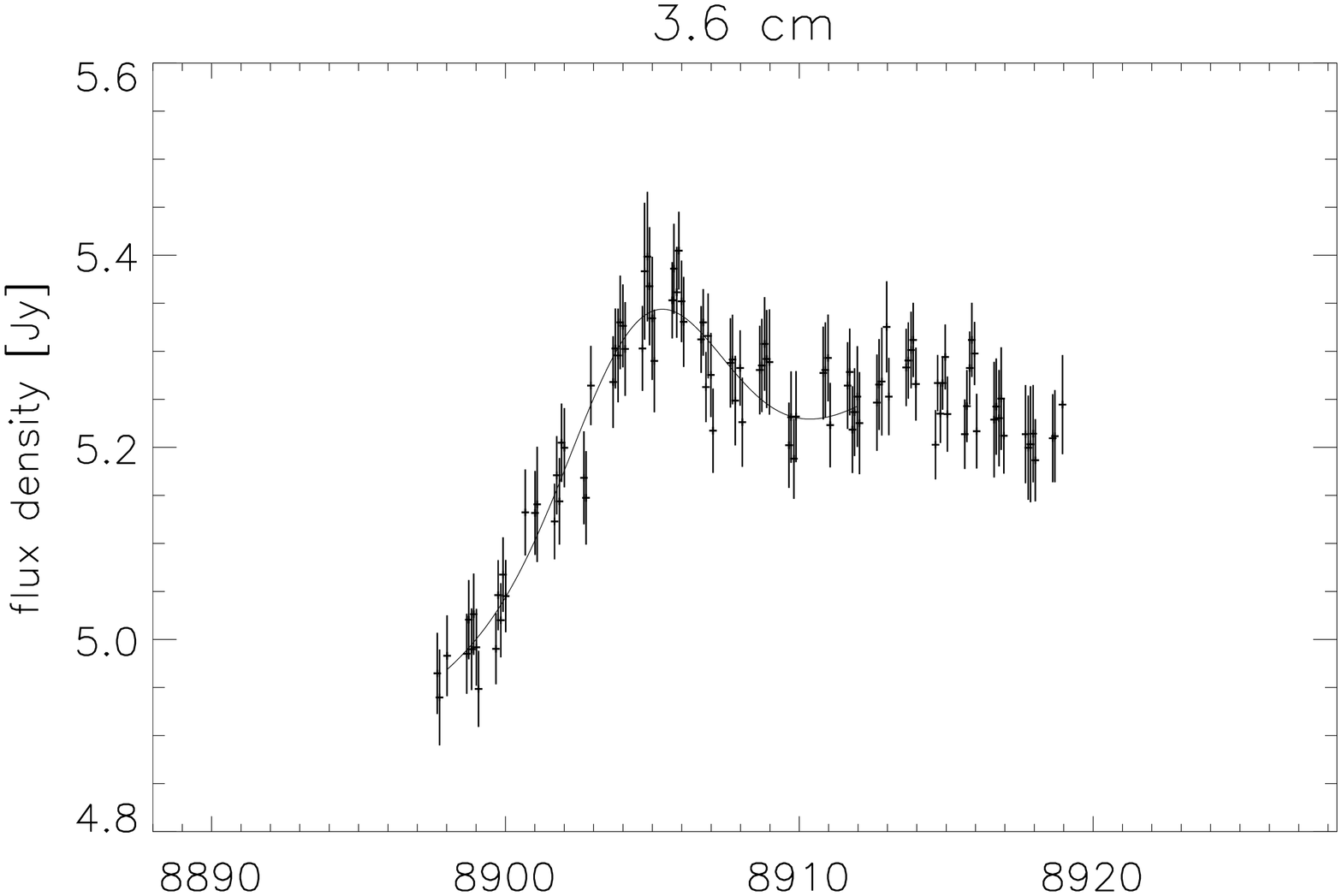,height=5cm,angle=90}
\psfig{figure=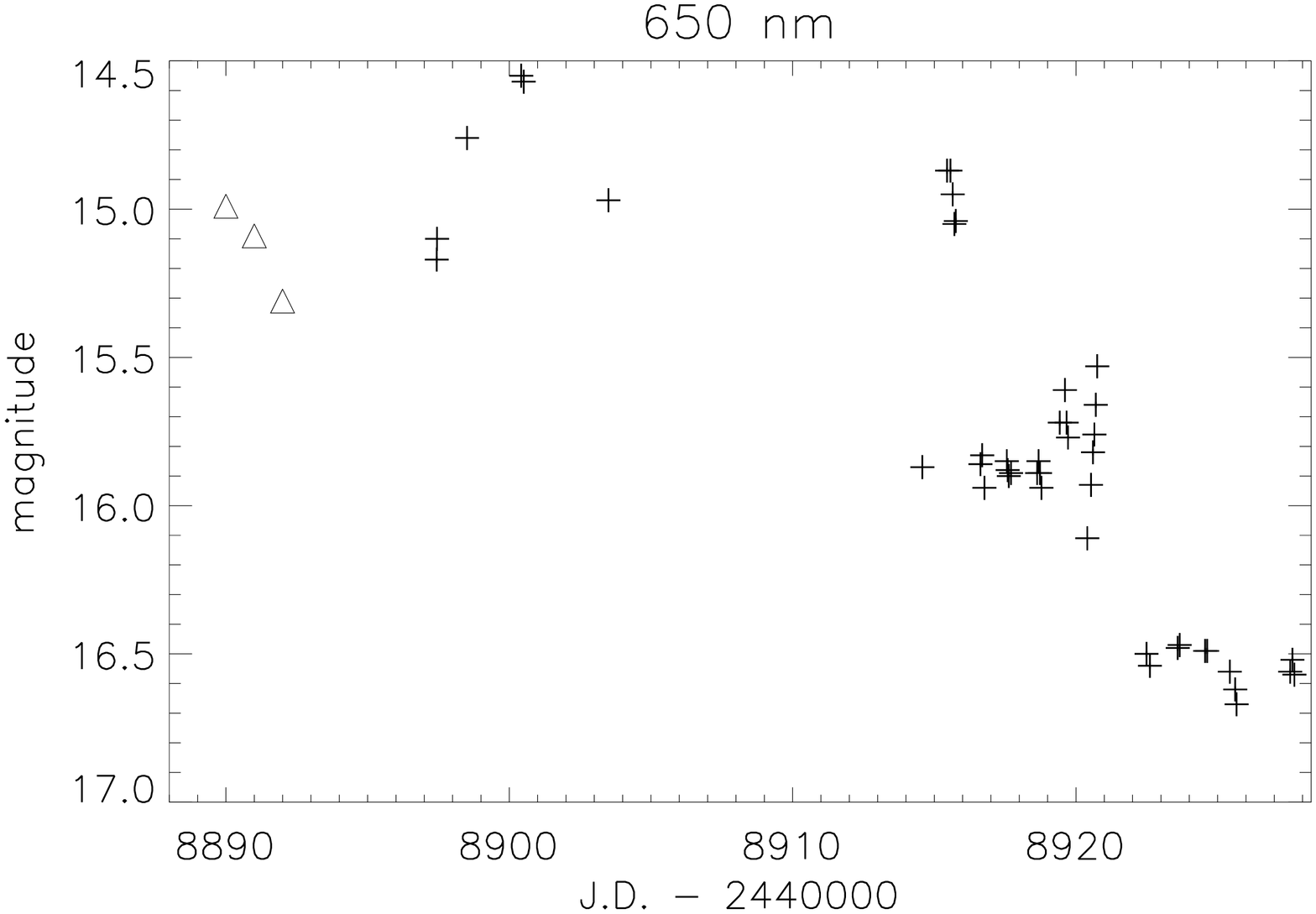,height=5cm,angle=90}
\caption{
\label{fg:mainplot}
Intensity variations of 0235+164 in Oct.\ 1992 at 1.49\,GHz
($\lambda=20$\,cm), 4.86\,GHz ($\lambda=6$\,cm), 8.44\,GHz ($\lambda
=3.6$\,cm), and in the optical R-band ($\lambda = 650$\,nm) (from top
to bottom). Plotted is the flux density (in Jy for the radio data, in
magnitudes for the optical data) versus Julian Date. For the radio
lightcurves Gaussians fitted according to Equation (1) are included
(see Section~2.3 and Table~\ref{gaussfit}). For the optical lightcurve
we included three data points (marked by triangles) measured by
Schramm et al.~(1994).}
\end{figure}

The resulting lightcurves for the three frequencies are displayed in
the top panels of Fig.~\ref{fg:mainplot}. The mean flux densities are
1.57\,Jy, 4.05\,Jy, and 5.22\,Jy for $\nu = 1.49, 4.86, 8.44$\,GHz,
respectively.  Therefore, 0235+164 had a highly inverted spectrum at
the time of the observations with spectral indices $\alpha^{1.5\,{\rm
Ghz}}_{4.9\,{\rm GHz}} = 0.80$ and $\alpha^{4.9\,{\rm GHz}}_{8.4\,{\rm
GHz}} = 0.46$.

\subsection{Optical observations}
\label{optical}

The radio data were supplemented by observations at 650\,nm (R-band
filters) taken at the following telescopes: 0.7\,m Telescope,
Landessternwarte Heidelberg, Germany; 1.2\,m Telescope, Observatoire
de Haute Provence, France; 1.2\,m Telescope, Calar Alto, Spain; 2.1\,m
Telescope, Cananea, Mexico.

Owing to limited observing time per source and weather limitations,
the optical data are sampled more sparsely. They cover the first week
of the radio observations, leave a gap for ten days, and continue for
a total of thirty days, thus ending ten days after the radio
monitoring. After the usual CCD reduction process (see e.g.\ Heidt \&
Wagner~1996), we performed relative photometry referencing the
measurements to three stars within the field. The corresponding
lightcurve is plotted in the bottom panel of Fig.~\ref{fg:mainplot}.
The measurement errors are smaller than the symbol size. In addition,
we include three data points, taken from the long-term monitoring by
Schramm et al.~(1994). Those are marked by triangles.

\subsection{Lightcurve analysis}

As evident from Fig.~\ref{fg:mainplot}, 0235+164 is variable in all
three radio bands and in the optical. A major flare around JD~2448905
can be identified throughout the radio frequencies, and may be
tentatively connected with the optical maximum at the beginning of the
observation.  We note, however, that the exact position of the latter
cannot be determined precisely due to the sparse sampling of the
optical data. Therefore, we consider the connection between radio and
optical variations as possible, but not definitive.

In addition, a second flare towards the end is present in the
21\,cm-data, possibly corresponding to the increase at 6\,cm, and the
sharp peak by a factor of two in the optical. A corresponding feature
at 3.6\,cm would be expected well inside the observation period but is
definitely not present.  The lightcurve at 6\,cm shows additional
faster variations which have no corresponding features at the other
wavelengths. These faster variations (which are not shown by the
calibrator sources and therefore are real) could for example be caused
by scattering in the ISM as we will discuss later. But we note that
the ``global'' behavior is very similar at all three radio
wavelengths.

We focus in this paper on the first flare (JD $\lessim$ 2448910) which
is pronounced in all three radio frequencies, and could be connected
to the optical increase around JD~2448900. We assume that all four
observed lightcurves are caused by the same physical event in the
source. In order to describe this major feature, we fit a linear
background and one Gaussian component to the radio lightcurves (using
all data points before JD~2448910) according to
\begin{equation}
S\,(t) = a_0 + a_1 \cdot t +\, a_2 \cdot
\exp\left(-\frac{(t-a_3)^2}{a_4^2}\right) \quad ,
\end{equation}
where S(t) is the measured flux density. The parameters and estimated
errors are listed in Table~\ref{gaussfit}.
\begin{table*}
\begin{center}
\begin{tabular}{|l|c|ccc|}
\hline
          & \multicolumn{1}{|c|}{Background} &
\multicolumn{3}{c|}{Gaussian flare}\\
$\lambda$ [cm] & \multicolumn{1}{|c|}{Slope}      &
\multicolumn{1}{c}{Amplitude} & \multicolumn{1}{c}{Center} &
\multicolumn{1}{c|}{Duration}\\
          & \multicolumn{1}{|c|}{$a_1$ [Jy/d]} & \multicolumn{1}{c}{$a_2$ [Jy]} &
\multicolumn{1}{c}{$a_3$ [JD]} & \multicolumn{1}{c|}{$a_4$ [d]}\\
\hline
20 & $(4.58 \pm 0.002) \cdot 10^{-3}$ & $0.318 \pm 0.01$
    & $8904.3 \pm 0.1$ & $1.67 \pm 0.1$  \\
6  & $(2.18 \pm 0.002) \cdot 10^{-2}$ & $0.299 \pm 0.01$
    & $8905.0 \pm 0.1$  & $2.44 \pm 0.2$  \\
3.6& $(1.99 \pm 0.003) \cdot 10^{-2}$ & $0.243 \pm 0.02$
    & $8904.8 \pm 0.3$  & $3.65 \pm 0.9$  \\
\hline
\end{tabular}
\end{center}
\caption[]{Fit to the radio lightcurves by a linear background
and one Gaussian component}
\label{gaussfit}
\end{table*}
The fits reveal three properties which make the radio variability
quite unusual. First, the relative amplitude of the flare becomes
larger with increasing wavelength. Second, the duration of the event
(i.e., the width of the Gaussian given by the parameter $a_4$)
decreases with increasing wavelength.  And third, no monotonic
wavelength dependence of the time of the peak can be found.  Including
the sparse optical data the sequence is rather: 650\,nm $\rightarrow$
20\,cm $\rightarrow$ 6/3.6\,cm (the peaks at 6\,cm and 3.6\,cm are
simultaneous within the errors).

We determine the time lags between the peaks by deriving Cross
Correlation Functions for the radio data sets (again using only data
before JD~2448910.0). The CCFs were computed using an interpolation
method (e.g.\ White \& Peterson \cite{white94} and references
therein). Afterwards, the time lags were determined by the calculation
of a weighted mean of the CCF (i.e., the center of mass point)
using all values  $\geq 0.5$. The resulting time lags are:
\begin{eqnarray}
\tau^{20\,{\rm cm}}_{6\,{\rm cm}} & = & 0.84 \, \mbox{days,} \nonumber \\
\tau^{20\,{\rm cm}}_{3.6\,{\rm cm}} & = & 0.71 \, \mbox{days,} \nonumber \\
\tau^{6\,{\rm cm}}_{3.6\,{\rm cm}} & = & -0.24 \, \mbox{days,} \nonumber
\end{eqnarray}
with an error of about 0.2 days in each case. The differences between
the time lags derived from the Gaussian fits and the CCF are within
the errors and probably due to the fact that the flares are not
perfectly Gaussian (this explains also that $\tau^{20\,{\rm
cm}}_{6\,{\rm cm}} + \tau^{6\,{\rm cm}}_{3.6\,{\rm cm}} \neq
\tau^{20\,{\rm cm}}_{3.6\,{\rm cm}}$). The deviation from the Gaussian
shape is particularly obvious in the lightcurve at 6\,cm.
Nevertheless, the CCF analysis confirms the result that the sequence
of the flares is unusual, since the 20\,cm peak clearly precedes the
peaks in the other bands, while the time lag between the 6 and the
3.6\,cm-data does not appear to be significant.

To check the significance of the time lags between the maxima, we
carried out Monte-Carlo-Simulations for the Cross-Correlations between
the radio frequencies. As a start model for the lightcurves we used
the Gaussian fit parameters with the original sampling and added
Gaussian noise by a random process. In a second step, we allowed the
sampling pattern to be shifted in time randomly and independently for
every single simulation. This procedure confirmed that the peak at
20\,cm significantly precedes the other two.

\subsection{Long-term variability}

\begin{figure}
\psfig{figure=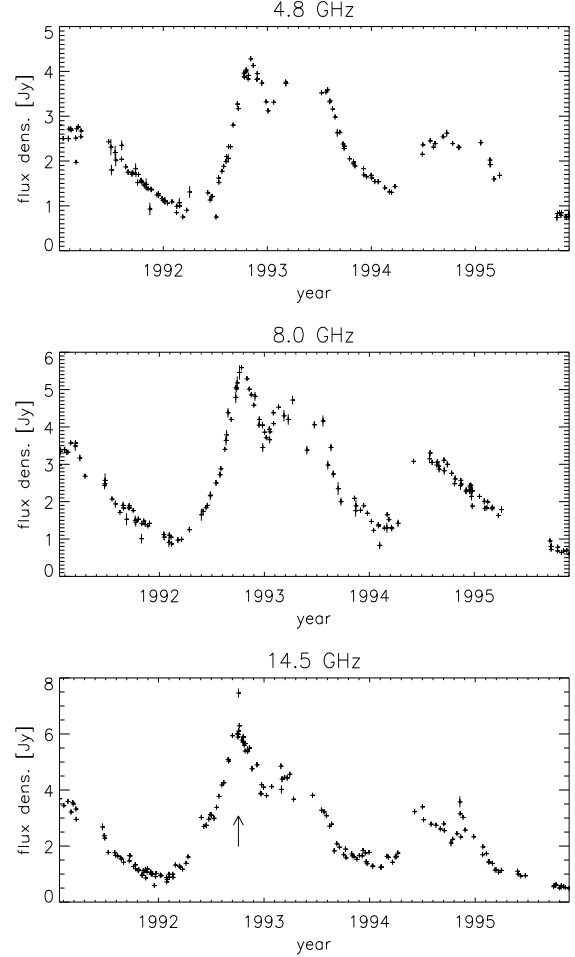,width=8.5cm,height=13cm}
\caption[]{Long-term monitoring of 0235+164 since 1991
with the UMRAO 26-meter telescope at 4.8\,GHz, 8\,GHz, and 14.5\,GHz (from top
to bottom). An arrow indicates the epoch of our VLA observations, close to the
peak of the large flux density outburst. }\label{fg:longplot}
\end{figure}

In Fig.~\ref{fg:longplot} we present radio data at 4.8, 8.0 and
14.5\,GHz obtained within the Michigan Monitoring Program (e.g.\ Aller
\cite{aller99} and references therein) from January 1991 to November
1995.  Our VLA observations (indicated by the arrow in the bottom
panel of Fig.\ \ref{fg:longplot}) coincide with the peak of a large
flux density outburst.

Also in the mm- and cm-radio data, published by Stevens et al.\
(\cite{stevens94}) a maximum at the time of the VLA-observations can
be seen at least at 22, 37, 90 and 150\,GHz. The optical data
presented by Schramm et al.\ (\cite{schramm94}) also give a clear
indication for an outburst at visible wavelengths right before our VLA
observations. Three data points which are close to our observations
are included in our R-band lightcurve (Fig.~\ref{fg:mainplot}, marked
as triangles.  The long-term monitoring implies that our observations
took place when 0235+164 was in a very bright state.

\section{Discussion}

\subsection{Problems}

On the basis of the collected data, and the analysis in the previous
section, we note several properties of the observed flare which are
unusual and require an explanation in the framework of physical
models:

\begin{itemize}

\item The sequence of the flares is rather unusual.  The 20\,cm
maximum precedes the maxima at 3.6\,cm and 6\,cm. The first optical
maximum -- if connected to the radio events -- is about four days
earlier.

\item The peaks become narrower and stronger with increasing radio
wavelength -- a unique behavior which is not seen in other sources and
not easily explained in any of the ``standard'' physical models.

\item In case of an intrinsic origin of the variability, one can
derive the corresponding source brightness temperature from the
duration of the event (e.g.\ Wagner \& Witzel~1995).  For $\lambda =
20$\,cm this yields $T_B \simeq 7 \cdot 10^{17}$\,K, far in excess of
the inverse Compton limit (Kellermann \& Pauliny-Toth
\cite{kellermann69}).

\item Our observations show that variations are present at both radio
and optical wavelengths, with very similar timescales. The gaps in the
optical lightcurve do not allow us to establish a one-to-one
correspondence between individual events in both wavelength ranges,
but it seems plausible that they are caused by a common physical
mechanism. This is a severe difficulty for models that attribute the
variations to strongly wavelength-dependent propagation effects
(free-free absorption and interstellar scintillation).

\end{itemize}

In the following, we discuss various models which could describe the
variations and take into consideration at least some of the peculiar
properties mentioned above.

\subsection{Relativistic shocks}

Propagation of a relativistic shock-front through the jet is commonly
accepted as one of possible causes of flux-density variability in AGN
(e.g.\ Blandford \& K\"onigl~1979, Marscher \& Gear~1985).  The time
scales usually involved in these models are of the order of weeks to
months (corresponding to source sizes in the range of light weeks to
months), and are consequently significantly longer than the ones
observed here.  Following Marscher \& Gear (\cite{marscher85}), the
characteristics of the flux density evolution in the case of a moving
shock within the jet can be described as follows. Starting at high
frequencies (in the sub-mm-regime) the outburst propagates to longer
wavelengths while the peak of the synchrotron spectrum shows a very
special path in the $S_{\rm m}$-$\nu_{\rm m}$-plane. This path can be
described by three power laws $S_{\rm m} \propto \nu_{\rm m}^k$ (with
different exponents $k$), distinguishing three different stages of the
evolution (see also Marscher \cite{marscher90}). During the
synchrotron or the adiabatic expansion stages, which are likely to be
found in this wavelength range, the spectral maximum is expected to
move from higher to lower frequencies, with the peak flux density
being either constant or decreasing with decreasing frequency. Thus,
for this ``standard-model'', we expect that the flux density reaches
its maximum at higher frequencies first, and that the amplitude of the
peak decreases towards lower frequencies.  This is contrary to our
observations.

In contrast, the canonical behavior for a shock-in-jet model is seen
in the long-term lightcurve (Fig.\ \ref{fg:longplot}): The amplitude
increases with increasing frequency, resulting in a strongly inverted
spectrum during the outburst and the sequence of the peaks
(determined from CCFs) follows the expectations: 14.5\, GHz
$\rightarrow$ 8.0\,GHz $\rightarrow$ 4.8\, GHz.

It should be noted, that the model of Marscher \& Gear (1985) is based
on three assumptions: (i) the instantaneous injection of relativistic
electrons, (ii) the assumption that the variable component is
optically thick at the beginning of the process, and (iii) that the
jet flow is adiabatic. Therefore, this model describes a transition
from large ($\tau > 1$) to small ($\tau < 1$) optical depths for each
frequency. It is possible, however, that 0235+164 is initially
optically thin at our observing wavelengths, and that the optical
depth increases with time, e.g.\ due to continuous injection of
electrons or field magnification, or through compression. In this
case, $\tau$ may reach unity -- and the flux density its maximum -- at
lower frequencies earlier than at higher ones (e.g.\ Qian et al.\
\cite{qian96}), as observed. A similar behavior was discussed for
CTA\,26 by Pacholczyk (\cite{pacholczyk77}) -- although on longer time
scales. In this model, we expect the maximum at 4.9\,GHz to precede
the one at 8.4\,GHz or that they are reached at the same time. The
latter may be true within the uncertainty. However, the different
amplitudes and the durations of the event cannot be explained without
additional assumptions.

Alternatively, the observed variations may be explained with a thin
sheet of relativistic electrons moving along magnetic field lines with
a very high Lorentz factor ($\gamma \simeq 20$--25). In this case, a
slight change of the viewing angle (e.g.\ from 0\deg\ to 2--3\deg) may
give rise to dramatic variations of the aberration angle and therefore
of the observed synchrotron emission (Qian et al., in preparation).
Additionally, this should cause significant changes in the linear
polarization (strength and position angle), which may be studied in
future observations.

\subsection{Precessing beam model}

We now investigate a scenario in which the observed effect is caused
by the variable Doppler boosting of an emitting region moving along a
curved three-dimensional path.  If the observed turnover frequency of
such a region falls between 1.5 and 8.4\,GHz, peaks in the lightcurves
can be displaced relative to each other. The Doppler factor variations
required to reproduce the observed timelags may be caused by a
perturbed relativistic beam (cf.\ Roland et al.\ \cite{roland94}, see
also Camenzind \& Krockenberger~1992). The jet is assumed to consist
of an ultra-relativistic ($\gamma\simeq 10$) beam surrounded by a
thermal outflow with speed $\beta\simeq 0.4$. The relativistic beam
precesses with period $P_0$ and opening angle $\Omega_0$. The period
of the precession may vary from a few seconds to hundreds of
days. Roland et al.\ (\cite{roland94}) show that this model can
explain the observed short-term variability of 3C\,273, and also makes
plausible predictions about the kinematics of superluminal features in
parsec-scale jets. We use a similar approach to describe the flux
evolution of 0235+164.  The trajectory of an emitting component inside
the relativistic beam is determined by collimation in the magnetic
field of the perturbed beam, and can be described by a helical
path. In the coordinate system (x,y,z) with z-axis coinciding with the
rotational axis of the helix, the component's position is given by
\begin{equation}
\label{helix:1}
\left\{
\begin{array}{rcl}
x & = & r(z) \cos(\omega t\,-\,k z\,+\, \phi_0) \\
y & = & r(z) \sin(\omega t\,-\,k z\,+\, \phi_0) \\
z & = & z(t),
\end{array}
\right.
\end{equation}
where $r(z)$ describes the amplitude of the helix, and can be
approximated as $r(z) = r_0 z / (a_0+z)$. For a precessing beam,
$a_0=r_0/\tan\Omega_0$. The form of the function $z(t)$ should be
determined from the evolution of the velocity $\beta_{\rm b}$ of the
relativistic component. $\beta_{\rm b}$ can be conveniently expressed
as a function of $z$, and in the simplest case assumed to be
constant. Then, under the condition of instantaneous acceleration of
the beam ($dz/dt>0$, for $z\rightarrow 0_+$), the component trajectory
is determined by
\begin{equation}
\label{helix:4}
t(z) = t_0 + \int^z_{z_0} \frac{1}{\dot{z}} dz = t_0 + \int^z_{z_{0}}
\frac{C_2(z)}{k\omega r^2(z) + C_3(z)} dz \quad ,
\end{equation}
with $C_1(z) = [r_z^{\prime}(z)]^2 + 1$, $C_2(z) = C_1(z) + k^2 r^2(z)$, and
$C_3(z) = [C_2(z) \beta^2_{\rm b} - \omega^2 r^2(z) C_1(z)]^{1/2}$. (This
follows directly from Equation (7) in Roland et al.\ (1994).) Generally, both
$\omega$ and $k$ can also vary. Their variations should be then represented
with respect to $z$, and $\omega(z)$ and $k(z)$ be used in Equation
(\ref{helix:4}).

\begin{figure}
\centerline{\psfig{figure=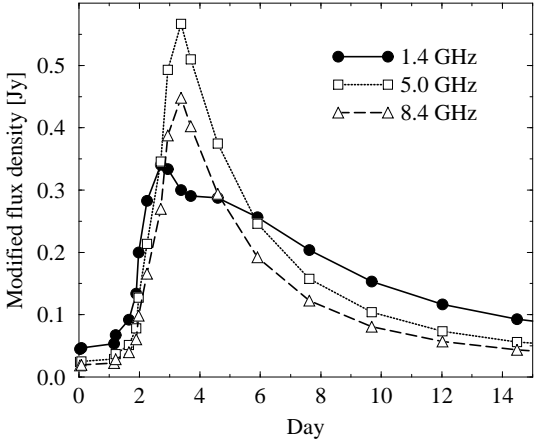,width=0.45\textwidth}}
\caption{Model lightcurves at 1.4, 4.9, and 8.4\,GHz for $P_0 = 200$\,days,
$\Omega_0 = 5\fdg7$. \label{fg:ltcurve}}
\end{figure}

We describe the emission of the perturbed beam by a homogeneous
synchrotron spectrum with spectral index $\alpha =-0.5$, and rest
frame turnover frequency $\nu_{\rm m}^{\prime} = 150$\,MHz. The beam
precession period $P_0 = 200$\,days, and $\Omega_0 = 5\fdg7$. $r(z)$
is described by $r_0 = 0.1$\,pc and $a_0 = 1$\,pc. The corresponding
lightcurves are plotted in Figure \ref{fg:ltcurve}.

We can see that the model is capable of reproducing the observed time
lag between 1.4\,GHz and the higher radio frequencies.  One can
speculate that a more complex physical setting (e.g.\ spectral
evolution of the underlying emission, or inhomogeneity of the emitting
plasma) may be required for explaining the apparent discrepancy
between the modeled and observed widths of the flare.

\subsection{Free-free absorption by a foreground medium}

Here we consider the effect of free-free absorption in a foreground
medium, either in the host of the BL\,Lac object itself, or in one of
the intervening redshift systems. To keep the discussion simple, we
neglect the cosmological redshift, i.e., factors $(1+z)$. The optical
depth for free-free absorption of a plasma is approximately given by
(see e.g.\ Lang \cite{lang74}):
\begin{equation}
\tau = 8.235 \cdot 10^{-2} \, T^{-1.35} \, \nu^{-2.1} \, \int N^2 \, dl \quad ,
\end{equation}
where $T$ is the electron temperature in K, $\nu$ is measured in GHz,
and the emission measure $\int N^2 \, dl$ in pc\,cm$^{-6}$. Thus, the
absorption of radiation by a foreground medium can be described by
$e^{-c \cdot \lambda^2}$, where $c$ is a constant. We assume the
following scenario. The source is moving with transverse speed $v$
behind a patchy foreground medium so that changes in the emission
measure towards the source produce variable absorption. To lowest
order, we describe gaps between the clouds by
\begin{equation}
\tau = k^2 \, \lambda^2 \, x^2
\end{equation}
($x$ being the axis perpendicular to the line of sight). Since the
observed flux density of a point source is given by $S_{\rm obs} = S
\, e^{-\tau}$, a moving source seen through such a gap in the
foreground medium will show peaked lightcurves with roughly Gaussian
shape. The width of the peaks will decrease with increasing
wavelength. However, there are two major problems that need to be
addressed. Firstly, in this model the maxima for all frequencies are
reached at the same time, and secondly, the observed durations (i.e.,
the widths of the Gaussians fitted according to Equation~(1)) do not
follow the expected behavior $a_4 \propto \lambda^{-1}$.

The time lags between the peaks of the observed lightcurves can be
explained for an extended source by a slight shift of the brightness
center depending on the frequency.

To deal with the second problem, we assume that the source is not
point-like, but has a circular Gaussian shape, with the source size
proportional to the wavelength.  Thus, the flux density is given by
\begin{equation}
S(x,t) = S_0 \cdot \exp \left(- \frac{(x-vt)^2}{\sigma_{\lambda}^2} \right)
\quad ,
\end{equation}
with $\sigma_{\lambda} = \sigma_0 \cdot \lambda$ (i.e., $\sigma_0$ is
the source size at 1\,m wavelength).

Assuming that the angular size of the variable region is much smaller
than the antenna beam, the observed flux density is given by the
integral
\begin{equation}
S_{\rm obs} (t) = \int^{+\infty}_{-\infty} S(x,t) \cdot e^{-\tau(x)} dx
\end{equation}

Evaluating the above integral, gives (since $S_{\rm obs}$ is the
integral of a product of two Gaussians)
\begin{equation}
S_{\rm obs} (t) = S' \cdot \exp
\left( \frac{-v^2t^2}{\sigma_{\lambda}^2 + 1 / k^2 \lambda^2} \right)
\end{equation}
with a new normalization constant $S'$. Therefore, the square of the width of
the Gaussian is
\begin{equation}
a_4^2 = \frac{1}{v^2} \left( \sigma_{\lambda}^2 + \frac{1}{k^2 \lambda^2}
\right)
= \frac{\sigma_0^2}{v^2}\lambda^2 +
\frac{1}{k^2 \, v^2}\frac{1}{\lambda^2} \quad ,
\label{eq:width}
\end{equation}
and it should depend on wavelength like $A \cdot \lambda^2 + B \cdot
\lambda^{-2}$. By adjusting the parameters $A$ and $B$ to fit
the measured values of $a_4^2$ at the three observing wavelengths, we
derive values for $\sigma_0^2/v^2$ and $k^2 \cdot v^2$. We assume here
that the transverse speed is dominated by superluminal motion with $v
/ c = \beta_{\rm app}$ and obtain a source size of $0.0067 \cdot
\beta_{\rm app}$\,pc corresponding to an angular size of $1.6 \cdot
\beta_{\rm app}\,\mu$as at $\lambda = 1$\,m. We note that such a small
source diameter even for $\beta_{\rm app} = 10$ results in a
brightness temperature of about $10^{15}$\,K, and therefore violates
the inverse Compton limit. For higher velocities -- as observed in
this source (e.g.\ Chu et al.\ \cite{chu96}) -- the observed size can
be larger. However, to reconcile our observational findings with the
inverse Compton limit, Doppler factors of the order of 100 are needed.

The second term in Equation (\ref{eq:width}) gives the size of the gap
in the foreground medium, i.e., the distance between the points where
$\tau = 1$.  Since we assumed $\tau = k^2 \, \lambda^2 \, x^2$, this
distance is $\Delta x = 2 / (k \cdot \lambda)$, which is about $2.3
\cdot 10^{-4} \cdot \beta_{\rm app}$\,pc at $\lambda = 1$\,m. (Note
that this is true only for the case where the absorber is at the
redshift of the BL\,Lac object; the ratio of the angular diameter
distances of emitter and screen has to be applied as a correction
factor in the case of an intervening absorber.)

We still have to check whether Equation (4) gives a sufficient optical
depth for reasonable choices of electron temperature and emission
measure. The strongest constraints come from the data at 3.6\,cm: to
explain the observed amplitude of 0.24\,Jy at a source flux of 5\,Jy,
$\tau$ must be at least 0.05 at this wavelength. For an electron
temperature of 5000\,K, an emission measure of $5 \cdot
10^6$\,pc\,cm$^{-6}$ is needed. The thickness of the absorber cannot
be much larger than the transverse scale derived above, which is
0.06\,pc at 3.6\,cm for $\beta_{\rm app} = 10$; this gives an electron
density of $10^4$\,cm$^{-3}$. These values are within the range found
in Galactic H\,II regions and planetary nebulae.

We conclude that this model can explain the observed shorter duration
of the flares at longer wavelengths, and -- under the assumption of
slightly different spatial locations of the brightness center at the
observed wavelengths -- also the sequence of the peaks. It predicts
that the amplitude of the peaks increases more strongly with
wavelength than observed, but it is consistent with the data when an
underlying non-variable component is taken into account. However, in
the possible case of a connection between the radio and the optical
variations this model fails, since the optical radiation would not be
affected by free-free-scattering.

\subsection{Interstellar scattering (ISS)}

Scattering processes in the interstellar medium are well known to
cause flux density variations at radio frequencies (e.g.\ Rickett
\cite{rickett90}). In this section we investigate the possibility that
ISS is the cause of the variations seen in our observations. We will
follow mainly the considerations and notations of Rickett et al.\
(\cite{rickett95}). For a point-like source, the spatial scale of flux
density variations caused by RISS is given by ($L$ is the path length
through the medium, $\theta_{\mbox{\scriptsize scat}}$ the scattering
angle)
\begin{equation}
r_{0,\lambda} \simeq 0.25 \, L \, \theta_{\mbox{\scriptsize scat}} \quad ,
\end{equation}
which is proportional to $\lambda^{2.2}$, for a Kolmogorov-type medium
(Rickett et al.\ \cite{rickett84}), and therefore also
$\theta_{\mbox{\scriptsize scat}} \propto \lambda^{2.2}$ (cf. Cordes et al.\
\cite{cordes84}).
The spatial scale of an extended source (assuming a Gaussian shape of
$\sigma_{\lambda}$ in width) is then given by
\begin{equation}
r_{\theta,\lambda} = \sqrt{r^2_{0,\lambda} + (0.5 \, L \, \sigma_{\lambda})^2}
\quad .
\end{equation}
Then, the scintillation index $m_{\theta,\lambda}$ and the variability
timescale $\tau_{\theta,\lambda}$ for the extended source
can be derived by
\begin{eqnarray}
m_{\theta,\lambda} & = &
        m_{0,\lambda} \frac{r_{0,\lambda}}{r_{\theta,\lambda}} \quad, \\
\tau_{\theta,\lambda} & = &
        \frac{r_{\theta,\lambda}}{V} \quad , \nonumber
\end{eqnarray}
where $V$ is the velocity of the Earth (i.e., the observer) relative
to the scattering medium, and $ m_{0,\lambda}$ is the (wavelength
dependent) scintillation index of a point source.

We assume a source diameter which is proportional to $\lambda$ as we
did in the previous section, thus $\sigma_{\lambda} = \sigma_0
\lambda$, and use $\theta_{\mbox{\scriptsize scat}} = \theta_0
\lambda^{2.2}$ (see above).  This gives
\begin{eqnarray}
m_{\theta,\lambda} & = & \frac{m_{0,\lambda} \, \theta_0 \, \lambda^{2.2}}
{ \sqrt{\theta_0^2 \, \lambda^{4.4} + 4 \, \sigma_0^2 \, \lambda^2} }
\quad \mbox{and}\\
\tau_{\theta,\lambda} & = & \frac{L}{4V}
\sqrt{\theta_0^2 \, \lambda^{4.4} + 4 \, \sigma_0^2 \, \lambda^2}
\quad . \nonumber
\end{eqnarray}
Therefore, it is clear that -- independent of the wavelength
dependence of $m_{0,\lambda}$ -- the timescales of the variations
become shorter for decreasing wavelengths. This is contrary to our
observational findings (see Table \ref{gaussfit}), implying that this
simple model is unlikely to explain the observations. Additionally,
interstellar scattering cannot cause variability in the optical
regime. Hence, in this case again, a possible connection of the
optical and the radio variations would rule out ISS as the only cause
of the observed variability.

However, owing to the small source diameters involved here, ISS can be
present as an additional effect. As an example, we calculate the
scintillation index and the timescales with the following
assumptions. Following Rickett (\cite{rickett86}) the path length in
the interstellar medium of our galaxy is $L \simeq 500\,{\rm pc} \cdot
\csc |b| \simeq 788\,{\rm pc}$ (the source galactic latitude is
$-40\deg$). With $\sigma_0 = 1.2$ mas (which corresponds to $T_{\rm B}
\simeq 10^{12}$\,K), $\theta_0 = 60$ mas and a typical velocity (of
the observer) $V = 50$\,km/s this yields:
\begin{center}
\begin{tabular}{|lcc|}
\hline
$\lambda$ [cm] & $m_{\theta,\lambda}$ & $\tau_{\theta,\lambda}$ [d] \\
\hline
20  &  0.48  & 12.2 \\
6 &  0.32  & 1.28 \\
3.6 & 0.21  & 0.64 \\
\hline
\end{tabular}
\end{center}
Therefore, the faster variations which are clearly seen at higher
frequencies (especially in the 6\,cm lightcurve) may be due to ISS.

\subsection{Gravitational Microlensing}

Another possible explanation for the origin of the observed variations
is gravitational microlensing (ML) by stars in a foreground galaxy. ML
effects have been unambiguously observed in the multiple QSO 2237+0305
(Irwin et al.\ \cite{irwin89}, Houde \& Racine \cite{houde94}), and
most likely also in other multiply-imaged QSOs (see Wambsganss
\cite{wambsganss93}, and references therein). The possibility that ML
can cause AGN variability has long been predicted (Paczy\'nski
\cite{paczynski86}, Kayser et al.\ \cite{kayser86}, Schneider \& Weiss
\cite{schneider87}), but it remains unclear whether ML causes a
substantial fraction of the observed variability in QSOs (e.g.\
Schneider \cite{schneider93}).

0235+164 has a foreground galaxy ($z=0.524$) situated within two
arcseconds from the line of sight (Spinrad \& Smith \cite{spinrad75}),
and an additional galaxy 0\farcs 5 away from the source (Stickel et
al.\ \cite{stickel88}, see also additional components reported in
Yanny et al.\ \cite{yanny89}).  Additionally, a nearby absorption
system was observed at $\lambda = 21$\,cm by Wolfe, Davis \& Briggs
(\cite{wolfe82}). All three objects may host microlenses affecting the
emission from 0235+164. Thus, for 0235+164 the probability for ML is
expected to be high (Narayan \& Schneider \cite{narayan90}), so that
sometimes ML events should be present in the lightcurves.

We will show now how ML can modulate the underlying long-term
lightcurve and explain faster variations of long-wavelength flux
compared to short-wavelength radiation, even when the longer
wavelength radiation comes from a larger source (component). Since the
available data do not permit a detailed account of possible ML
situations, the attention here is restricted to two simple situations:
an isolated point-mass lens in the deflector, and a cusp singularity,
formed by an ensemble of microlenses (Schneider \& Weiss
\cite{schneider87}, Wambsganss \cite{wambsganss90}). In fact, both
cases yield similar predicted ML lightcurves. The scales of the source
size and the lens mass necessary to yield a flux variation of the
observed kind can be estimated for both cases together.

\begin{figure}\centering
\mbox{\psfig{figure=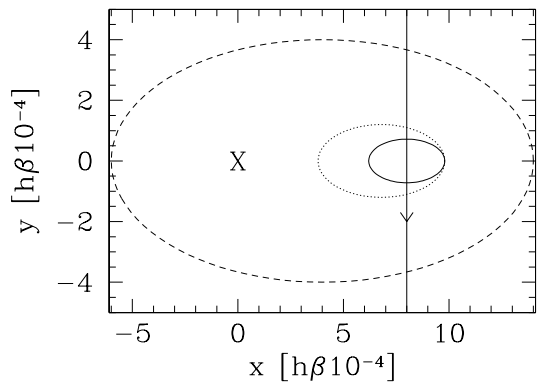,width=0.48\textwidth}}
\caption{Geometry of the proposed microlensing scenario. Elliptical
surface brightness contours are drawn for the 3.6\ cm (solid), 6\ cm
(dotted), and 20\ cm (dashed) source components. The caustic point of
the microlens is at the origin, denoted by `X', and the source
components are assumed to move along the solid line as indicated. The
scales indicated on the axis correspond to the synthetic lightcurves
shown in Fig.\ \ref{fg:ml_lightcurves}. }
\label{fg:ml_geometry}
\end{figure}

We assume an elliptically shaped emitting feature that moves
relativistically in the direction roughly coinciding with the minor
axis of the ellipse. Such a component can be formed by relativistic
electrons which are locally accelerated by a shock front inside a
superluminal jet. The shape of the source component and its
orientation is then determined by the flow inside the jet. A Gaussian
brightness profile is assumed, with component size $\propto \lambda$
(see Fig.\ \ref{fg:ml_geometry} for details). We postulate that the
emission peaks at all three wavelengths are displaced relative to each
other, but that the peaks of shorter-wavelength components are
situated within the half intensity contour of longer-wavelength
components.

Let $\beta c$ be the apparent effective transverse velocity of the
source component; using the redshift $z_s=0.94$ of the object, this
corresponds to an angular velocity of $v_{\rm a}=2\beta h\,
10^{-4}$\,mas/day. If a source component moves along a track in the
source plane, and the component size is much smaller than the minimum
angular separation $d_{\rm a}$ from the singularity, as indicated in
Fig.\ \ref{fg:ml_geometry} (solid ellipse), then the timescale of
variation is given roughly by the ratio $d_{\rm a}/v_{\rm a}$.  On the
other hand, if a strongly elongated source component moves so that
parts of it cross the line of sight to the singularity (as indicated
in Fig.\ \ref{fg:ml_geometry}, dashed ellipse), then the shortest
possible timescale is roughly the ratio between the transverse angular
source size $a_\lambda$ (the minor semi-axis at wavelength $\lambda$)
and the angular velocity. Now assume that the former case approximates
the 3.6\,cm source and the latter case approximates the 20\,cm
source. If $\Delta t_{3.6}\sim 4$\,days, $\Delta t_6\sim 3$\,days and
$\Delta t_{20} \sim 2$\,days are the variability timescales for the
three wavelengths considered, we have
\begin{equation}
d_{\rm a}\sim v_{\rm a}\,\Delta t_{3.6}\sim 8\beta h \, 10^{-4}{\,\rm
mas}\quad ,
\end{equation}
and
\begin{equation}
r \, a_{20}\sim 4 \beta h \, 10^{-4}\;{\rm mas}\quad ,
\end{equation}
where $r \leq 1$ is the axis ratio of the Gaussian source
component. In order for the 20\,cm source to experience appreciable
variations, the closest separation of its center from the singularity
cannot be larger than its major semi-axis, i.e., $a_{20}\gtrsim d_{\rm
a}$, and this inequality can be satisfied for $r\lessim 0.5$.

Since the relative contribution of the moving component to the total
flux of the source is unknown, we cannot use the observed lightcurves
to determine the magnification of the component emission. The
magnification of a point source at separation $\theta$ from the point
singularity is
\begin{equation}
\mu_{\rm p}={x^2+2\over x\sqrt{x^2+4}}\quad,
\end{equation}
where $x=\theta/\theta_0$, and $\theta_0$ is the angular scale induced
by a point mass lens of mass $M$:
\begin{equation}
\theta_0=\sqrt{{4GM\over c^2}{D_{\rm ds}\over D_{\rm s} D_{\rm d}}} \quad,
\end{equation}
where $D_{\rm d}$, $D_{\rm s}$, and $D_{\rm ds}$ denote, respectively,
the angular diameter distances to the lens, the source, and from the
lens to the source, $m=M/M_\odot$ is the lens mass in units of the
solar mass. Assuming that the lens is situated at $z=0.525$,
\begin{equation}
\theta_0=1.87 \sqrt{m h}\,10^{-3}\;{\rm mas}\quad .
\end{equation}
Approximating the point-source magnification by $\mu_{\rm p}\simeq
1/x$ (for $x \ll 1$), and assuming as before that the size $a_{3.6}$
is much smaller than the closest separation of the source from the
point-like singularity, the maximum magnification of this source
component becomes
\begin{equation}
\mu_{\rm 3.6,max}\sim{\theta_0\over d_{\rm a}}
\simeq 2.34\,m^{1/2}\,h^{-1/2}\,\beta^{-1}\quad .
\end{equation}
Hence, a solar-mass star would yield a magnification of the order of 2
for the smallest source component moving at roughly the speed of light
and in general can produce lightcurves similar to the observed
variations.

\begin{figure}\centering
\mbox{\psfig{figure=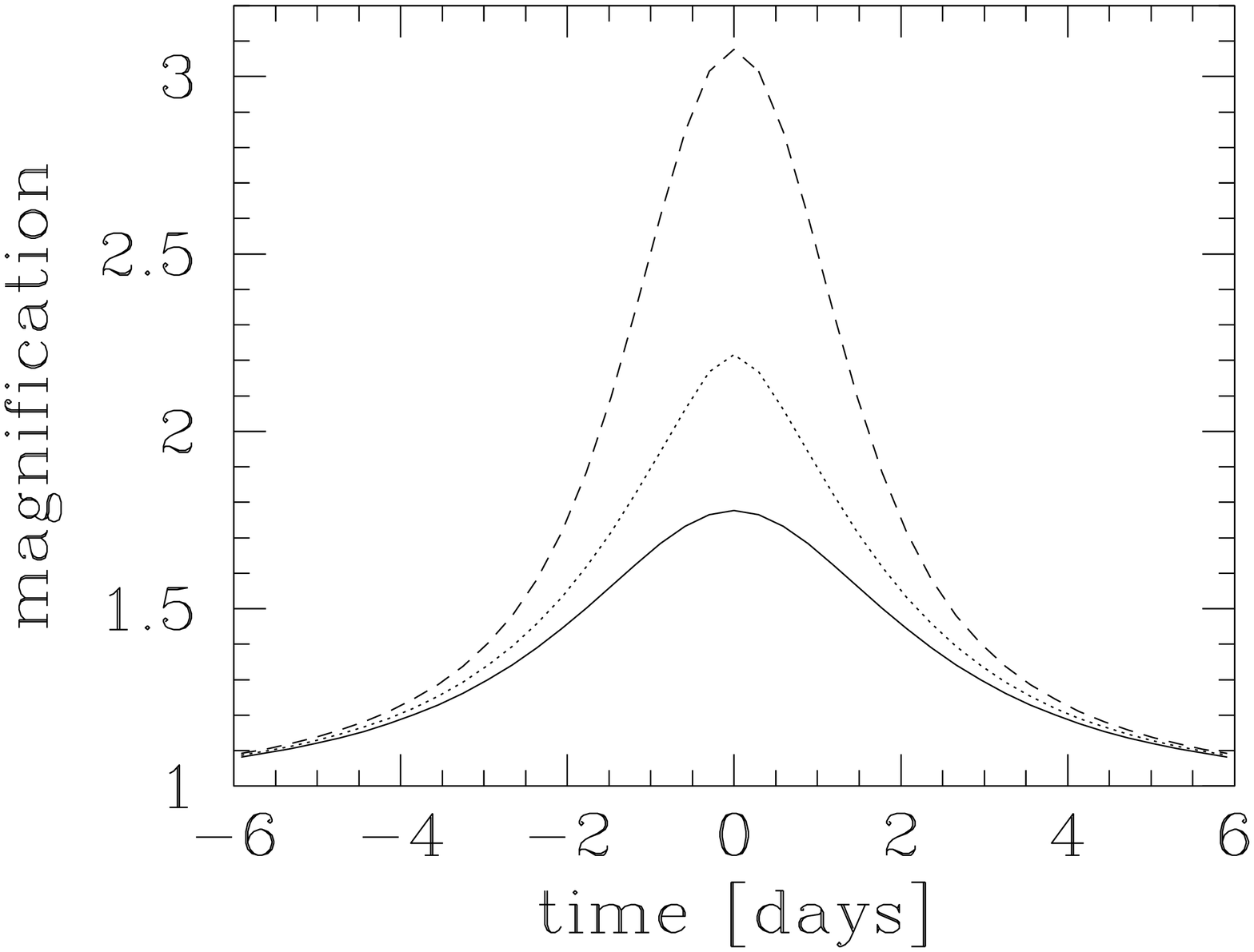,width=0.48\textwidth}}
\caption{Microlensing lightcurves obtained from the model discussed in
the text. The three curves correspond to the 3.6\,cm (solid), 6\,cm
(dotted) and 20\,cm (dashed) source components.}
\label{fg:ml_lightcurves}
\end{figure}

In Fig.~\ref{fg:ml_lightcurves}, we plot numerically determined ML
light\-curves for a moving source with an axis ratio $r=0.4$, minimum
separation $d_{\rm a}=8\beta h 10^{-4}$\,mas, and semi-major axis of
the 20\,cm source component of $a_{20}=\beta h 10^{-3}$\,mas. The lens
mass is $m=0.4 \beta^2 h$. The source sizes are chosen to be
proportional to wavelength, and the brightness peaks of the 6\,cm and
20\,cm components are displaced relative to the peak of the 3.6\,cm
component by 0.4 of their corresponding sizes. As can be seen from the
modeled lightcurves, the variability timescale of the 20\,cm component
is considerably shorter than that of the shorter wavelength
components, in accordance with our analytical estimates. In addition,
the observed shift of the brightness peak at 20\,cm before those at
smaller wavelengths can be accounted for in our model by a slight tilt
of the direction of motion of the source relative to the minor axis of
the surface brightness ellipses, in the sense of the large component
crossing the caustic point before the closest approach of the 3.6\,cm
component to that point. Nevertheless, we note that the small source
sizes needed (in the range of $\mu$as) will result in brightness
temperatures of the order of $10^{15}$\,K, i.e., three orders of
magnitude above the inverse Compton limit.
A more detailed modeling of the lightcurves by a microlensing
scenario is not warranted at this stage, given the large number of
degrees of freedom.  Nevertheless, the above considerations have
demonstrated that the basic qualitative features can be understood in
the microlensing picture without very specific assumptions.

\section{Conclusions}

We have observed the BL\,Lac object 0235+164 at three radio
wavelengths and in the optical R-band and found rapid variations in
all frequency bands.  One single event that can be identified at all
radio wavelengths shows very peculiar properties. The brightness peak
is reached first at 20\,cm wavelength, and afterwards at 3.6 and
6\,cm. The amplitudes of the flares decrease from longer to shorter
radio wavelengths, and the timescales become longer. The event in the
radio regime might be connected to the bright peak in the optical
lightcurve, although this connection remains questionable due to the
sparse sampling of the R-band data. In the previous sections, we have
discussed some models and to what extent they can explain the observed
variations.

While the conventional application of the shock-in-jet model has
difficulties in reproducing the observations, the assumption of an
increasing optical depth (e.g.\ due to continuous injection of
relativistic electrons) can cause a delay of the maximum at high
frequencies with respect to the lower frequencies, and therefore
explain at least one of the special features.

Variable Doppler boosting can cause simultaneous short-term
variability in all observed wave bands. Fairly pronounced time lags
between the different frequencies can be caused by turnover frequency
variations in the observed spectrum of a moving source. However,
broader peaks are expected at longer wavelengths.

Free-free absorption and interstellar scattering are only capable of
explaining radio variations, not variability in the optical
regime. Therefore, if the connection between the optical and the radio
variability is real, these models are ruled out as the only cause for
the variations. Furthermore, the dependence of the timescales on
wavelength argues against an explanation of the flare by interstellar
scattering. The absorption by a patchy foreground medium can easily
describe the shape and the widths of the flares (in the radio) and can
-- if we assume different locations for the brightness center -- also
explain the time sequence of the brightness peaks.

Gravitational Microlensing -- in combination with a
wavelength-dependent source size and a slight displacement of the
brightness peak -- provides a possible explanation for the observed
variations in the radio regime. One would also expect fairly strong
variability in the visible range, because of the much smaller source
size. Microlensing thus appears to be a viable explanation of the
observations, which is also quite attractive because of the known
foreground objects.

It is quite remarkable that these attempts to explain the rapid
radio variability in 0235+164 -- different as they are -- all imply
that the intrinsic source size is very small. To reconcile the
observations with the 10$^{12}$\,K inverse Compton limit, a Doppler
factor substantially higher than the ``canonical'' value of 10 (see
e.g.\ Ghisellini et al.\ 1993, Zensus~1997) is required.  Most
scenarios that we have investigated imply ${\cal D} \simeq 100$. In
this context it is interesting to note that circumstantial evidence
for superluminal motion with $\beta_{\rm app} \sim 30$ has been found
in this source (Chu et al.\ 1996). The variations in 0235+164 are also
among the strongest and fastest of all sources in the Michigan
monitoring program (e.g.\ Hughes et al.\ 1992).  This suggests that
the distribution of Doppler factors in compact radio sources has a
tail extending to ${\cal D} \simeq 100$, and that 0235+164 -- and
perhaps more generally the sources showing strong intraday radio
variability -- belong to this tail. The implied extremely small source
size can allow rapid intrinsic variations, and at the same time favor
propagation effects. It is therefore plausible that the observed
variability is caused by a superposition of both mechanisms.

\acknowledgements{ We thank I. Pauliny-Toth and E. Ros for critically
reading the manuscript, the referee, J.R. Mattox, for valuable
comments, C.E.\ Naundorf and R.\ Wegner for help with the
observations, and B.J.\ Rickett for stimulating discussions. The
National Radio Astronomy Observatory is a facility of the National
Science Foundation, operated under a cooperative agreement by
Associated Universities, Inc. This research has made use of data from
the University of Michigan Radio Astronomy Observatory which is
supported by the National Science Foundation and by funds from the
University of Michigan. }

\end{document}